\documentclass[twocolumn,5p,times]{article}
\usepackage{graphicx}
\usepackage{lscape,graphicx}
\usepackage{hhline}
\usepackage{multicol}
\usepackage{colordvi}
\usepackage{dcolumn}
\usepackage[fleqn]{amsmath}
\usepackage[mathscr]{euscript}
\usepackage[usenames, dvipsnames]{color}
\usepackage[usenames,dvipsnames,svgnames,table]{xcolor}
\usepackage[linktocpage]{hyperref}
\usepackage[a4paper,bindingoffset=0.2in,left=0.5in,right=0.5in,top=1.0in,bottom=1.0in,footskip=0.25in]{geometry}
\newcommand*{\doi}[1]{\href{http://dx.doi.org/#1}{doi: #1}}
\usepackage[numbers]{natbib}
\makeatletter
\renewcommand{\section}{\@startsection{section}{1}{\z@}{-3.5ex \@plus -1ex \@minus -.2ex}{1.3ex \@plus.2ex}{\normalfont\small\bfseries\boldmath}}
\makeatother

\makeatletter
\renewcommand{\subsection}{\@startsection{subsection}{2}{\z@}{-3.5ex \@plus -1ex \@minus -.2ex}{1.3ex \@plus.2ex}{\normalfont\small\bfseries\boldmath}}
\makeatother

\makeatletter
\renewcommand{\subsubsection}{\@startsection{subsubsection}{3}{\z@}{-3.5ex \@plus -1ex \@minus -.2ex}{1.3ex \@plus.2ex}{\normalfont\small\bfseries\boldmath}}
\makeatother

\makeatletter
\everymath{\if@display\else\thickmuskip=0mu plus 0mu\fi}
\makeatother

\title{\large \bf {\tt ee$\in$MC}: Arbitrary Initial Spin States for the Production of $\bf e^{+}e^{-} \to \tau^{+}\tau^{-} (\gamma) $ Events and the Impact on Spin Correlations.}
\date{}

\author{\normalsize Ian M. Nugent$^{*}$ \\ \normalsize Victoria, B.C., Canada}

\begin{document}
\twocolumn[
  \begin{@twocolumnfalse}
    \maketitle
\begin{abstract}
We present a modified spin algorithm, including spin correlations, which has been implemented in {\tt ee$\in$MC} for the simulation of $e^{+}e^{-} \to \tau^{+}\tau^{-} (\gamma)$ 
events with an arbitrary initial spin configuration. This algorithm is suitable for the proposed BELLE-II polarization upgrade to SuperKEKB \cite{USBELLEIIGROUP:2022qro,Roney:2021pwz}, 
both for the ideal case and for the case where
there are misalignments, radiative effects and allows for changing polarization conditions. The spin $1/2$ states of the polarized beams are constructed in terms of a super-positioning of the 
helicity states corresponding to the polarimetric-vector describing the overall beam polarization and implemented both through a modification of the initial state in the modified 
Altarelli-Parisi Density Function as well as a change of basis of the helicity states.  
\end{abstract}
\end{@twocolumnfalse}
]

\renewcommand{\thefootnote}{\fnsymbol{footnote}}
\footnotetext[1]{Corresponding Author \\ \indent   \ \ {\it Email:} inugent.physics@outlook.com}
\renewcommand{\thefootnote}{\arabic{footnote}}
\section{Introduction}

Recent proposals for an upgrade to SuperKEKB for polarization measurements at the BELLE-II experiment using a polarized $e^{-}$ beam are motivated by 
precision electro-weak measurements of the $sin^{2}\theta_{W}$ using the right-left asymmetry from pair production of $e^{+}e^{-}$, $\mu^{+}\mu^{-}$, $\tau^{+}\tau^{-}$, $c\bar{c}$ and $b\bar{b}$, 
measurements of the Michel Parameters, Tau electric-dipole-moment, Tau g-2 and the potential enhancements of new physics models \cite{USBELLEIIGROUP:2022qro,Roney:2021pwz}.   
In the proposed upgrade to the SuperKEK design \cite{USBELLEIIGROUP:2022qro}, the $e^{-}$ beam circles the ring with a polarization vector along the vertical (y) axis, 
orthogonal to the KEK-ring plane, and is only rotated longitudinally along the $e^{-}$ beam direction near the interaction point (IP). There are three proposed designs for rotating 
the spin axis: the BINP Spin Rotator Concept; Dipole-Solenoid-Quadrupoles combined function Magnets; and Direct Wind Magnets for the Spin Rotators \cite{USBELLEIIGROUP:2022qro}. 
Within these proposed design concepts, the longitudinal polarization is expected to reach $\sim 70-80\%$, however, the transverse components are not necessarily eliminated at the 
interaction point \cite{USBELLEIIGROUP:2022qro}. 
To evaluate the impact of the spin dynamic for polarized beams on experimental measurements, is it essential to be able incorporate the initial state spin dynamics for a 
broad range of accelerator and detector conditions, this includes mis-alignments, spin relaxation distributions and radiative effect, into the
simulation for the determination of the efficiency and for evaluating the impact on physics observables in a given analysis.
An algorithm which satisfies these criteria and which can be modified on an event-by-event biases is described in the 
following sections.

\section{Overview of Generator Formalism}

{\tt ee$\in$MC} \cite{Nugent:2022ayu}, is a stand-alone MC Generator which includes the random-number generation \cite{MT32,MT64Tab,MT64F2,xorshift,Knuth:1981}, the phase-space 
generation and the theoretical models for the QED processes $e^{-}e^{+}\to\mu^{+}\mu^{-}(n\gamma)$ and $e^{-}e^{+}\to\tau^{+}\tau^{-}(n\gamma)$ and decays of the $\tau$ lepton.
The phase-space generators are extensions of the  algorithm from \cite{Byckling:1969} modified to include embedded importance sampling \cite{Lopes:2006,Gelman:2014} constructed in terms of the 
structure of the relevant physics models to optimize the simulation.
The QED cross-section for $e^{-}e^{+}\to\mu^{+}\mu^{-}(n\gamma)$ and $e^{-}e^{+}\to\tau^{+}\tau^{-}(n\gamma)$ is determined within the Yennie-Frautschi-Suura Exponentiation 
Formalism \cite{Yennie:1961},

\begin{equation}
\resizebox{0.375\textwidth}{!}{$
d\sigma= \frac{\sum_{n=0}^{\infty} Y_{i}(Q_{i}^{2})Y_{f}(Q_{f}^{2})|\sum_{k=1}^{\infty}\bar{{\mathcal M_{a}^{b}}}|^{2} dPS_{a}^{\delta M}}{4(|\vec{P}_{e^{-}}|E_{e^{+}}+E_{e^{-}}|\vec{P}_{e^{+}}|)}
$}
\label{eq:YSR}
\end{equation}

\noindent  where the matrix elements are computed directly from the Feynman Diagrams using an object-orientated formalism \cite{Nugent:2022ayu}. 
$Y_{i/f}(Q_{i}^{2})$ represents the initial-state/final-state Yennie-Frautschi-Suura Exponentiation which has been implemented for: the  Yennie-Frautschi-Suura  calculation \cite{Yennie:1961}; 
the {\tt KK2F} approximation  \cite{kk2f}; Sudakov Form-Factor \cite{Peskin:1995ev}; and the Full LO calculation from \cite{Schwinger:1998} applying corrections from 
\cite{Nugent:2022ayu,Peskin:1995ev,Smith1994117} for the Coulomb potential. Renormalization is incorporated into the running of electromagnetic coupling constant  \cite{Jegerlehner,Sturm_2013} 
by means of Wards Identity \cite{Mandl:1985bg}. Details for the renormalization and Exponentiation in {\tt ee$\in$MC} can be found in \cite{Nugent:2022ayu}. 
The $\tau$ decays are simulated for both leptonic, $\tau^{-}\to l^{-}\bar{\nu}_{l}\nu_{\tau}$\footnote{The Charge conjugate is implied throughout this paper.},  and
semi-leptonic decays, $\tau^{-}\to h_{x} \nu_{\tau}$, at  Born level. The hadronic currents are implemented in the context of several theoretical frame-works: 
Chiral-Resonance-Lagrangian Models \cite{Finkemeier:1995sr,Kuhn:1990ad,Jadach:1993hs,Decker:1992kj};  
``Chromoelectric Flux-Tube Breaking Model''  within a ``revitalized [${^{3}P_{0}}$] quark model(s)''\cite{Isgur:1988vm,Godfrey:1985xj,Kokoski:1985is,Isgur:1983wj,Isgur:1984bm};
Vector-Dominance Models  \cite{BONDAR2002139,Bondar:1999}; and phenomenological models  \cite{CLEO3pi,Edwards:1999fj,Feindt:1990ev}.

\section{Initial State Polarization and the Spin Algorithm}

The beam polarizations can be characterized in terms of the polarimetric vector ($\vec{\mathbf{P}}$) representing the average spin 
direction of the interacting initial state electron or positron in the center-of-mass (CM) reference frame. The magnitude $|\vec{\mathbf{P}}|$ represents the strength of polarization or 
probability of the electron or positron being in a polarized state. In terms of the quantum mechanical spin-projection operator, the expectation value for the polarization vector may be written as

\begin{equation}
\resizebox{0.400\textwidth}{!}{$
\begin{array}{lll}
<S_{\hat{\mathbf{n}}}>&=&<s_{1/2,m}| \hat{\mathbf{s}}_{\hat{\mathbf{n}}} |s_{1/2,m}> \\&=&<s_{1/2,m}| \frac{\hbar}{2}\left(n_{x}\mathbf{\hat{\sigma}_{x}}+n_{y}\mathbf{\hat{\sigma}_{y}}+n_{z}\mathbf{\hat{\sigma}_{z}}\right) |s_{1/2,m}>
\label{eq:Spinorbraket}
\end{array}
$}
\end{equation}

\noindent where $\hat{\mathbf{n}}$ is a unit vector defined by $\vec{\mathbf{P}}=|\vec{\mathbf{P}}|\times \hat{\mathbf{n}}$. The wave-function of the electron and positron corresponding to a spin
aligned along the polarimetric vector can be constructed by means of a linear super-positioning of the longitudinal helicity states using the two orthogonal eigen-vectors of the spin-projection operators
$\hat{\mathbf{S}}_{\hat{\mathbf{n}}}=\frac{\hbar}{2} \left(n_{x}\mathbf{ \hat{\sigma}_{x}}+n_{y}\mathbf{\hat{\sigma}_{y}}+n_{z}\mathbf{\hat{\sigma}_{z}}\right)$
corresponding to the polarimetric vector. The generalized eigen-vector with $+1 (a_{ \hat{\mathbf{n}}})$ and $-1 (b_{ \hat{\mathbf{n}}})$ eigen-values may be written as

\begin{equation}
a_{ \hat{\mathbf{n}}}=
\left(\begin{array}{l}
cos(\frac{\theta}{2}) \\
sin(\frac{\theta}{2})e^{\imath\phi}
\end{array} \right),\ \ 
b_{ \hat{\mathbf{n}}}=
\left(\begin{array}{l}
sin(\frac{\theta}{2}) \\
-cos(\frac{\theta}{2})e^{\imath\phi}
\end{array} \right)
\end{equation}
  
\noindent respectively. In the context of the modified Altarelli-Parisi Density Function, from which the spin algorithms in {\tt ee$\in$MC} are based,
the linear super-position of states is incorporated in the spin sum through the projection operators \cite{Mandl:1985bg,Jadach:1984}. Adapting the initial state of the modified Altarelli-Parisi 
Density Function from 
\cite{Jadach:1984,Collins:1987cp,Knowles:1988hu} to include the Pauli-spin operator corresponding to the polarimetric vectors 
yields

\begin{equation}
\resizebox{0.4\textwidth}{!}{$
\begin{array}{ll}
\rho_{\lambda_{i},\lambda_{i^{\prime}}}^{(e^{-})}\rho_{\lambda_{j},\lambda_{j^{\prime}}}^{(e^{+})}\times  \mathcal{M}_{a}^{b}{_{\lambda_{i},\lambda_{j}\lambda_{k}\lambda_{l},...,\lambda_{n}}^{}}\mathcal{M}_{a}^{b}{_{\lambda_{i^{\prime}},\lambda_{j^{\prime}}\lambda_{k^{\prime}}\lambda_{l^{\prime}},...,\lambda_{n}}^{*}}
\times \prod\limits_{\alpha=k}^{n} D_{\lambda_{\alpha},\lambda_{\alpha}^{\prime}}
\label{eq:qedpolarAP}
\end{array}
$}
\end{equation}

\noindent where $\rho_{\lambda_{i},\lambda_{i^{\prime}}}^{(e^{-})}$ ($\rho_{\lambda_{j},\lambda_{j^{\prime}}}^{(e^{+})})$ are related to the $SU(2)$ representation of polarimetric vector \cite{Jadach:1984} 
corresponding to the $e^{-}$ ($e^{+}$) and are constructed from the weighted eigen-vectors for the $SU(2)$ Pauli-spin operators. The representation employed for the polarimetric vector 
of the initial state leptons is

\begin{equation}
\vec{\mathbf{P}} = F_{p}\hat{\mathbf{n}}=
\left(\begin{array}{l}
F_{p}\sqrt{1-(P_{l}/F_{p})^{2}}cos(\phi_{\hat{\mathbf{n}}}) \\
F_{p}\sqrt{1-(P_{l}/F_{p})^{2}}sin(\phi_{\hat{\mathbf{n}}}) \\   
P_{l}    
\end{array} \right) 
\end{equation}

\noindent
and is defined in terms of two parameters, the $\phi_{\hat{\mathbf{n}}}$ of the lepton in the transverse ($x-y$) plane\footnote{Note, the definition for the angle $\phi_{\hat{\mathbf{n}}}$ 
for both the $e^{-}$ and $e^{-}$ is in the CM 
frame coordinates where the $P_{l}$ is defined along relative to the momentum of the $e^{-}$ ($e^{+}$), the  $\hat{z}$  ($-\hat{z}$) axis.} and polarization fraction ($F_{p}$), in addition to the 
longitudinal polarization ($P_{l}=cos(\theta)\times F_{p}$). The $F_{p}$ allows for the inclusion of both unpolarized contributions when defining an arbitrary spin vector for the
super-positioning of the spinor states and must necessarily be $P_{l}\le F_{p} \le 1$. $P_{l}$, $\phi_{\hat{\mathbf{n}}}$ and $F_{p}$ can be defined on an event by event basis for 
both the initial $e^{-}$ and $e^{+}$ 
leptons allowing for a distribution of spin states. This allows for the inclusion of changing polarization conditions, in particular tails caused by mis-alignment, spin relaxation distributions
or radiative effects.   
From the latter form of the modified Altarelli-Parisi Density Function, the corresponding normalized spin probability is defined as

\begin{equation}
\resizebox{0.4\textwidth}{!}{$
\begin{array}{ll}
  P_{a}^{b} =  \rho_{\lambda_{i},\lambda_{i^{\prime}}}^{(e^{-})}\rho_{\lambda_{j},\lambda_{j^{\prime}}}^{(e^{+})} \times \frac{ \mathcal{M}_{a}^{b}{_{\lambda_{i},\lambda_{j}\lambda_{k}\lambda_{l},...,\lambda_{n}}^{}}\mathcal{M}_{a}^{b}{_{\lambda_{i^{\prime}},\lambda_{j^{\prime}}\lambda_{k^{\prime}}\lambda_{l^{\prime}},...,\lambda_{n}}^{*}}}
{|\bar{\mathcal{M}}|^{2}}
\times \prod\limits_{\alpha=k}^{n} D_{\lambda_{\alpha},\lambda_{\alpha}^{\prime}}
\label{eq:ProbAP}
\end{array}
$}
\end{equation}

\noindent 
where following the procedure from \cite{Nugent:2022ayu}, the normalized real spin probability density function is determined by applying the completeness-relation and 
projection operators to the outgoing $\tau^{-}/\tau^{+}$ states and decay matrices \cite{Mandl:1985bg,Jadach:1984}. Alternatively, a change of basis can be applied to the 
initial state spinors by directly constructing the spinor states out of a super-positioning of the helicity states using a weighting of the orthogonal eigen-vectors for the spin-projection operator 
corresponding to the polarimetric vector. The algorithms in \cite{Nugent:2022ayu}, were extended to include both the modified Altarelli-Parisi Density Function extended to include 
the $SU(2)$ representation of polarimetric vector for the $e^{-}e^{+}$ pair, as well as, the direct super-positioning of the helicity states\footnote{The two approaches are included in the {\it ee$\in$MC} as
a means of verifying the numerical implementation. Both methods are theoretically equivalent.}. 
The longitudinal factorized approach is modified to: 

\begin{enumerate}
\item  The given $e^{+}e^{-}\to\ l^{+}l^{-}(n\gamma)$ scattering process at order $n$, including the exponentiation procedure \cite{Nugent:2022ayu}, is simulated using the spin-average summed matrix element, 
\begin{equation}
\resizebox{0.400\textwidth}{!}{$
\begin{array}{ll}
|\mathcal{M}_{a}^{b}|^{2}=\rho_{\lambda_{i},\lambda_{i^{\prime}}}^{(e^{-})}\rho_{\lambda_{j},\lambda_{j^{\prime}}}^{(e^{+})}\times  \mathcal{M}_{a}^{b}{_{\lambda_{i},\lambda_{j}\lambda_{k}\lambda_{l},...,\lambda_{n}}^{}}\mathcal{M}_{a}^{b}{_{\lambda_{i^{\prime}},\lambda_{j^{\prime}}\lambda_{k^{\prime}}\lambda_{l^{\prime}},...,\lambda_{n}}^{*}},
\end{array}
$}
\end{equation} 
where indices $a$ and $b$ represent the number of radiated hard photons and internal photon lines respectively. 
For the super-positioning case, the helicity state indices in the spin sum
corresponds to the change of basis into the eigen-vector of the spin-operator $\hat{\mathbf{S}}_{\hat{\mathbf{n}}}$ where  $\lambda_{i}=\lambda_{i^{\prime}}$ and $\lambda_{j}=\lambda_{j^{\prime}}$.
\item  The probability distribution $P_{\lambda_{k},\lambda_{l}}$ is determined by contracting only the initial state indices in the modified Altarelli-Parisi Density Function \ref{eq:qedpolarAP}, 
where $\lambda_{k}=\lambda_{k}^{\prime}$ and $\lambda_{l}=\lambda_{l}^{\prime}$. 
An accept/reject algorithm is then applied using $P_{\lambda_{k},\lambda_{l}}$ to select the final state helicities with which the $\tau$ decays are simulated.
The probability distribution $P_{\lambda_{k},\lambda_{l}}$ is computed in the longitudinal basis for the outgoing leptons for both cases 
and is independent of the initial state basis applied in the spin sum. 
\item Once the event is simulated, the transverse spin correlations are then included by means of a second accept/reject algorithm using the probability from the 
modified Altarelli-Parisi Density Function adapted to include the transverse initial-state spin effects, Eq. \ref{eq:ProbAP}, which is bounded by $0\le P\le 4$. 
At each step in the accept/reject algorithm, a random number is used for each $\tau$ lepton
to rotate the $\tau$ decay products about the longitudinal axis of the respective $\tau$, and the probability from the modified Altarelli-Parisi Density Function based on Eq. \ref{eq:ProbAP}, 
is recomputed.

\end{enumerate}

\noindent In the non-factorized approach:

\begin{enumerate}
\item  The given $e^{+}e^{-}\to\ l^{+}l^{-}(n\gamma)$ scattering process at order $n$, including the exponentiation procedure \cite{Nugent:2022ayu}, is simulated for the spin-average summed matrix element
using the same procedure as the factorized approach. 
\item The outgoing $\tau$ leptons are then decayed in an unpolarized state.
\item The spin structure is implemented in the simulate using an accept/reject algorithm based on the probability from the modified Altarelli-Parisi Density Function in Eq. \ref{eq:ProbAP}. 
The final state decay products for each $\tau$ decay are rotated by a random $\phi$ and $\theta$ in the CM frame of respective $\tau$ decay before recomputing the probability from 
the modified Altarelli-Parisi Density Function adapted to include the transverse spin components of the initial state.   
\end{enumerate}

\section{Impact on the Differential Cross-section and Correlation in $\tau$ Decays}

\begin{figure*}[tbp]
\begin{center}
  \resizebox{520pt}{185pt}{
    \includegraphics{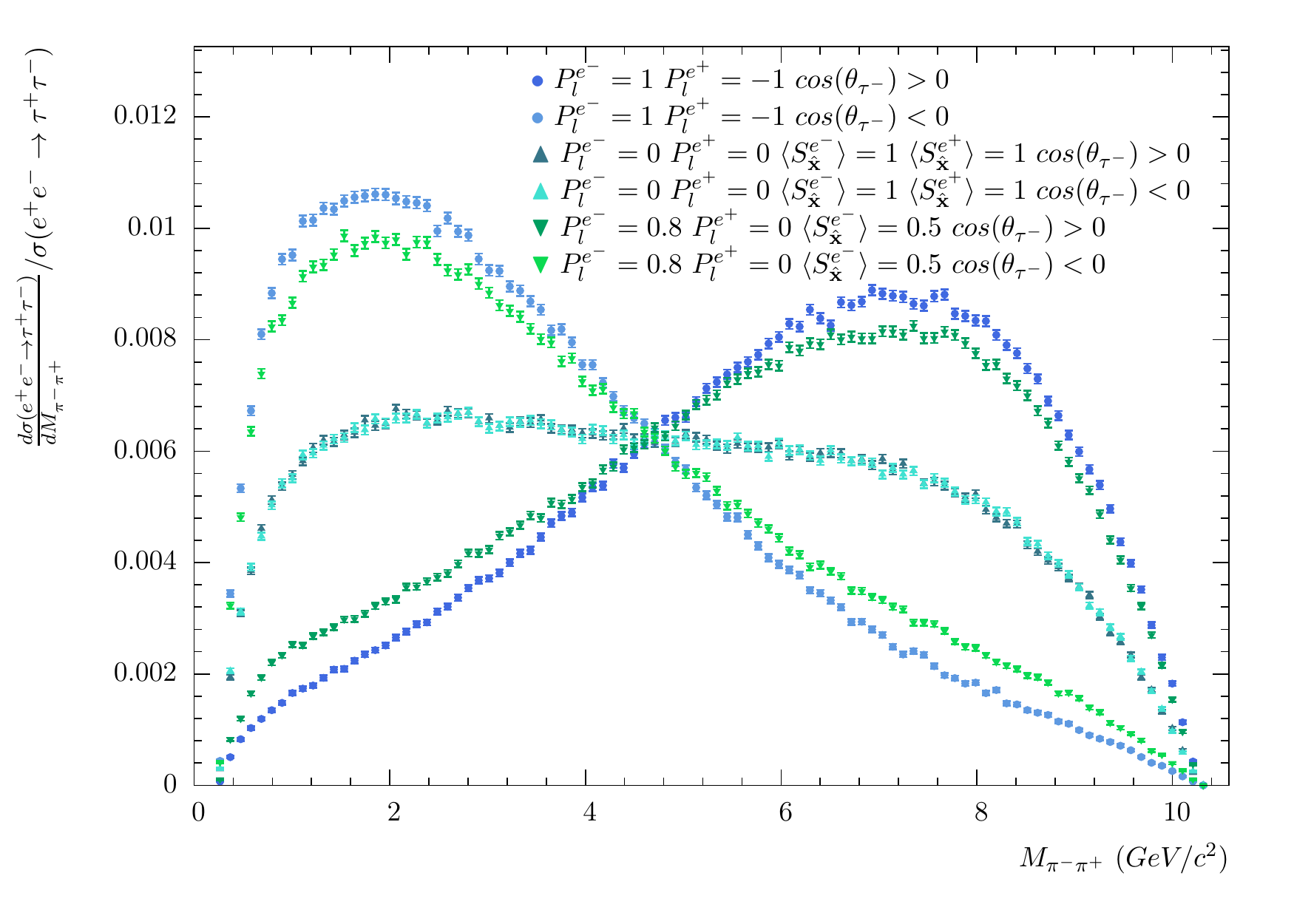}
    \includegraphics{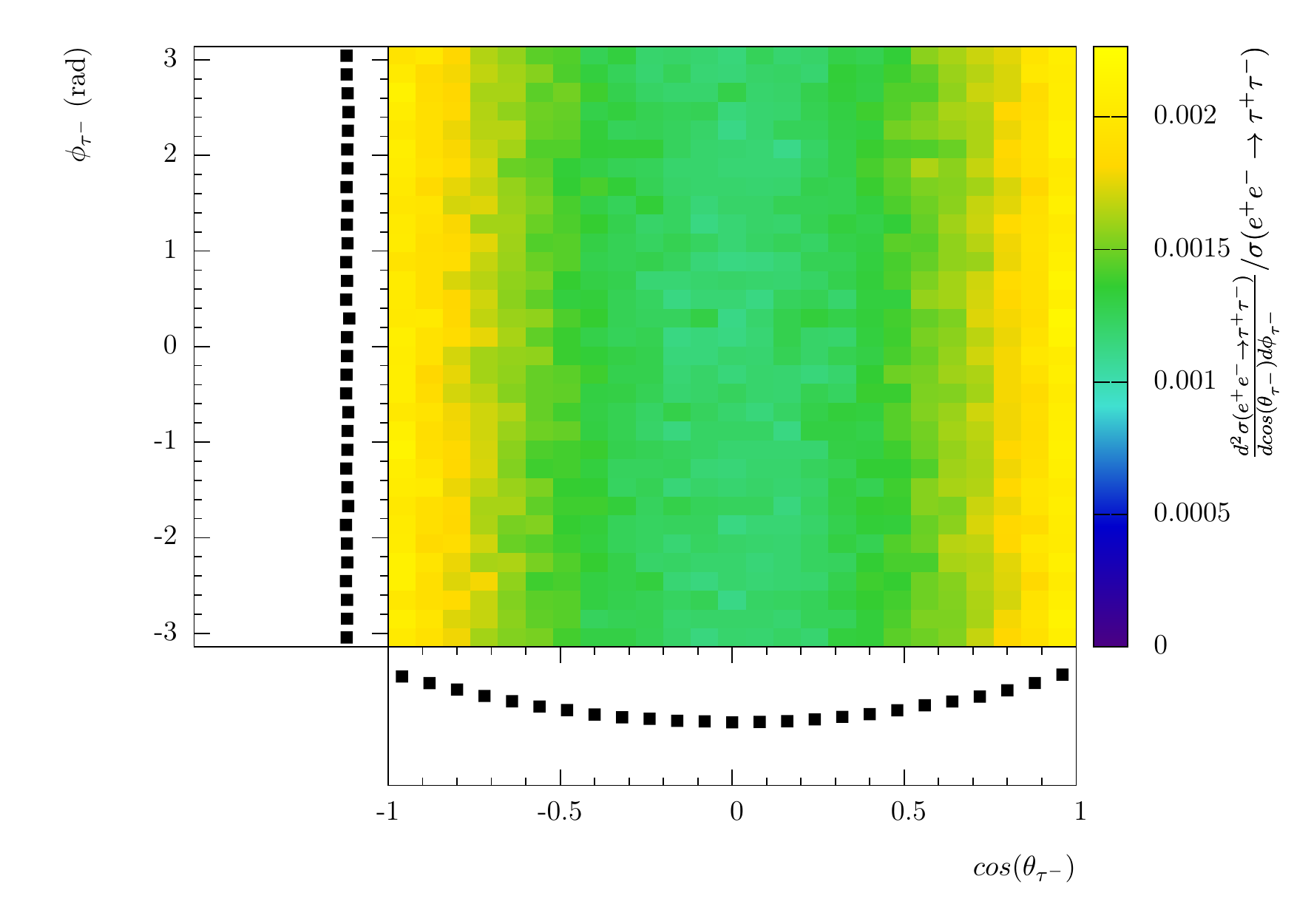}
  }
\resizebox{520pt}{185pt}{
    \includegraphics{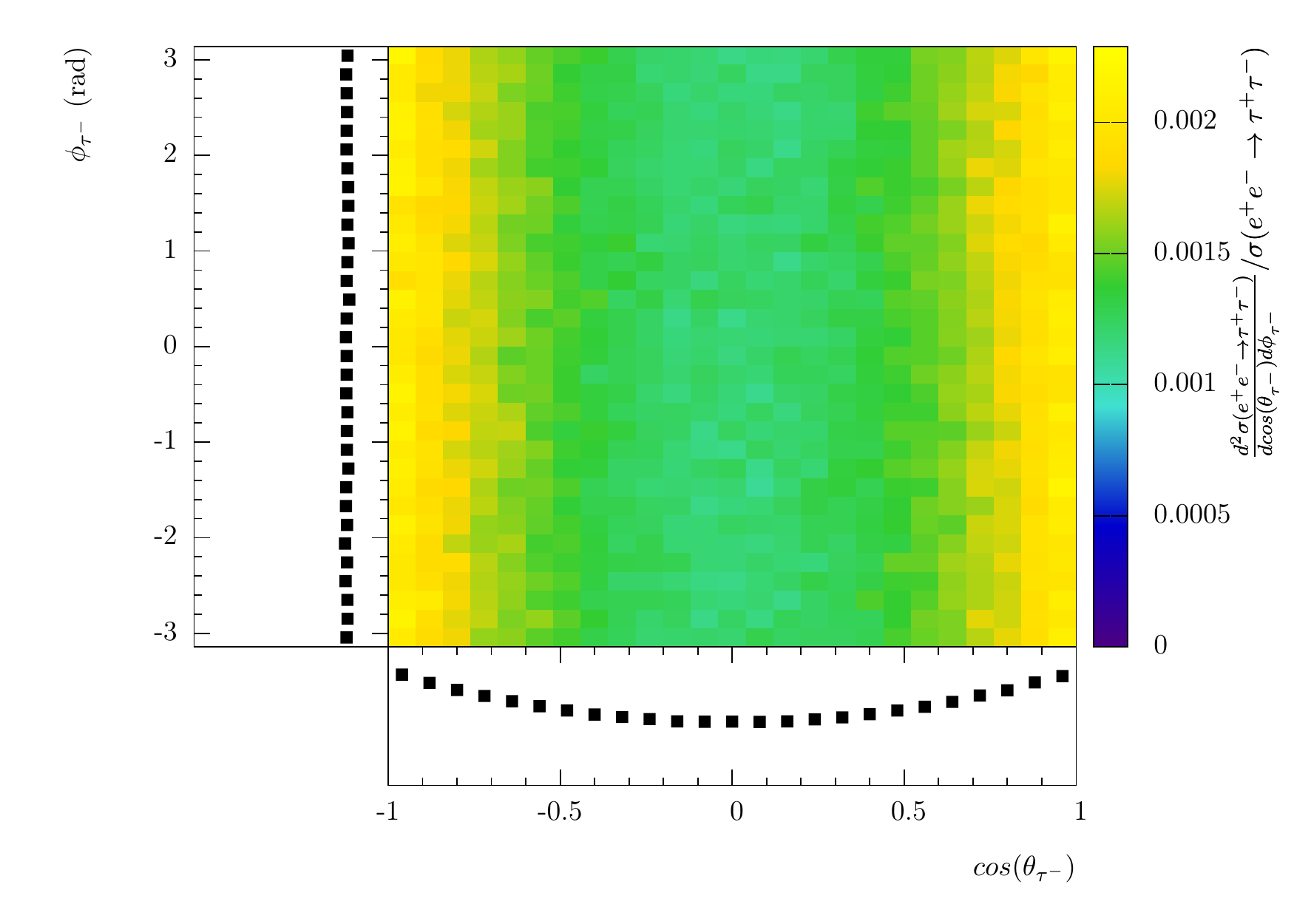}
    \includegraphics{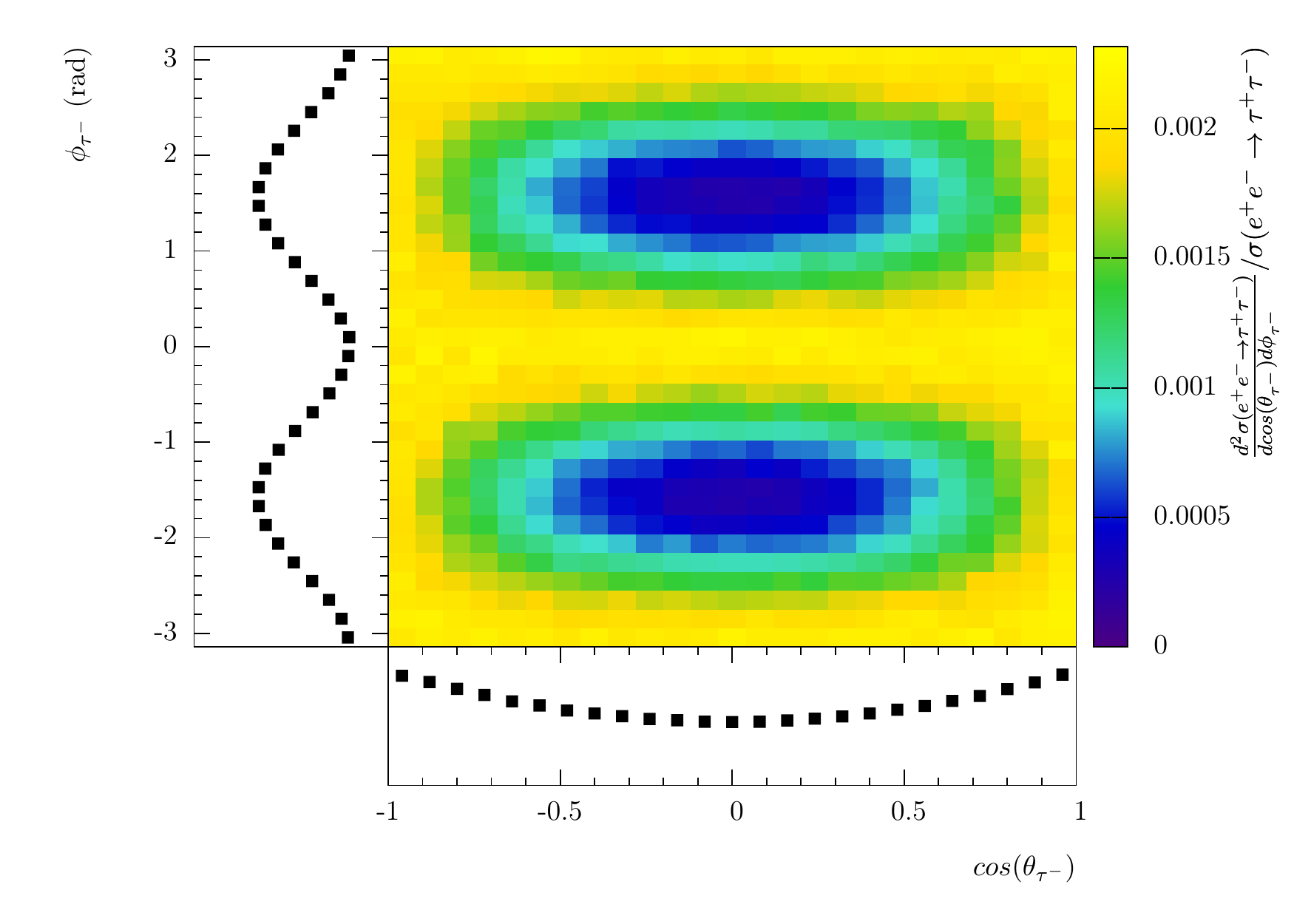}
  }
  \caption{The longitudinal observable $M_{\pi\pi}$  $cos(\theta_{\tau^{-}})>0$ and $cos(\theta_{\tau^{-}})<0$ distributions (upper-left) for longitudinally polarized beams, 
for a transversely polarized beam, 
and for a realistic polarization scenario at the proposed polarization upgrade at Belle-II \cite{USBELLEIIGROUP:2022qro}. There is a strong dependence between the
$cos(\theta_{\tau^{-}})>0$ and $cos(\theta_{\tau^{-}})<0$ distributions for the longitudinally polarized sample, while the transversely polarized sample is independent of
$cos(\theta_{\tau^{-}})>0$ and $cos(\theta_{\tau^{-}})<0$.   The cross-section as a function $cos(\theta_{\tau^{-}})$ and $\phi_{\tau^{-}}$ for a 100\% longitudinal 
polarized $e^{+}$ and $e^{-}$ sample (lower-left) and for a 100\% transversely polarized (along the x-axis) $e^{+}$ and $e^{-}$  sample (lower-right) 
and for a realistic polarization scenario at the proposed polarization 
upgrade at Belle-II \cite{USBELLEIIGROUP:2022qro} (upper-right). To simplify the
interpretation of these diagrams, they have been simulated at Born level.  \label{fig:LPolar}}
\end{center}
\end{figure*}

\begin{figure*}[tbp]
\begin{center}
\resizebox{520pt}{185pt}{
    \includegraphics{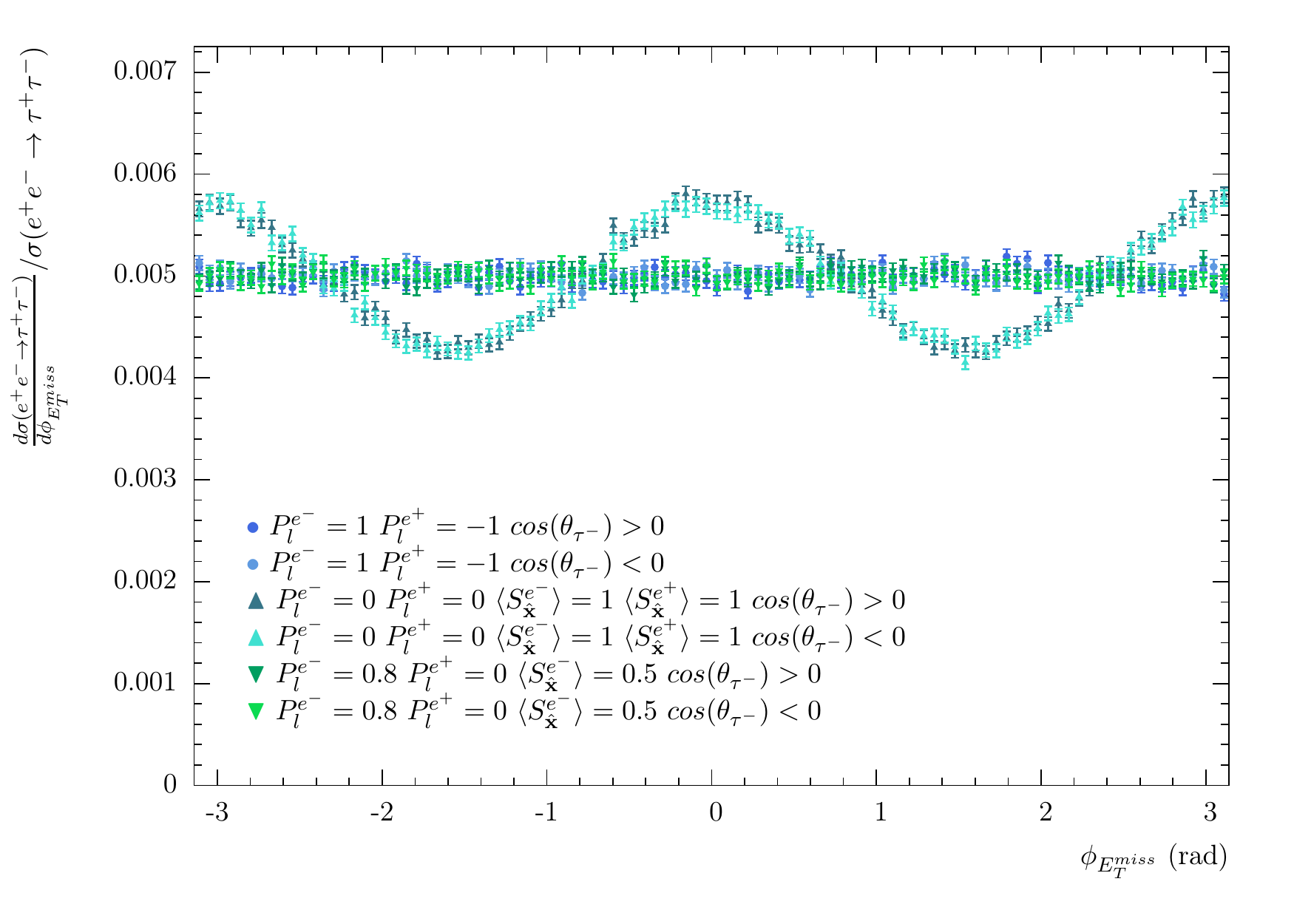}
    \includegraphics{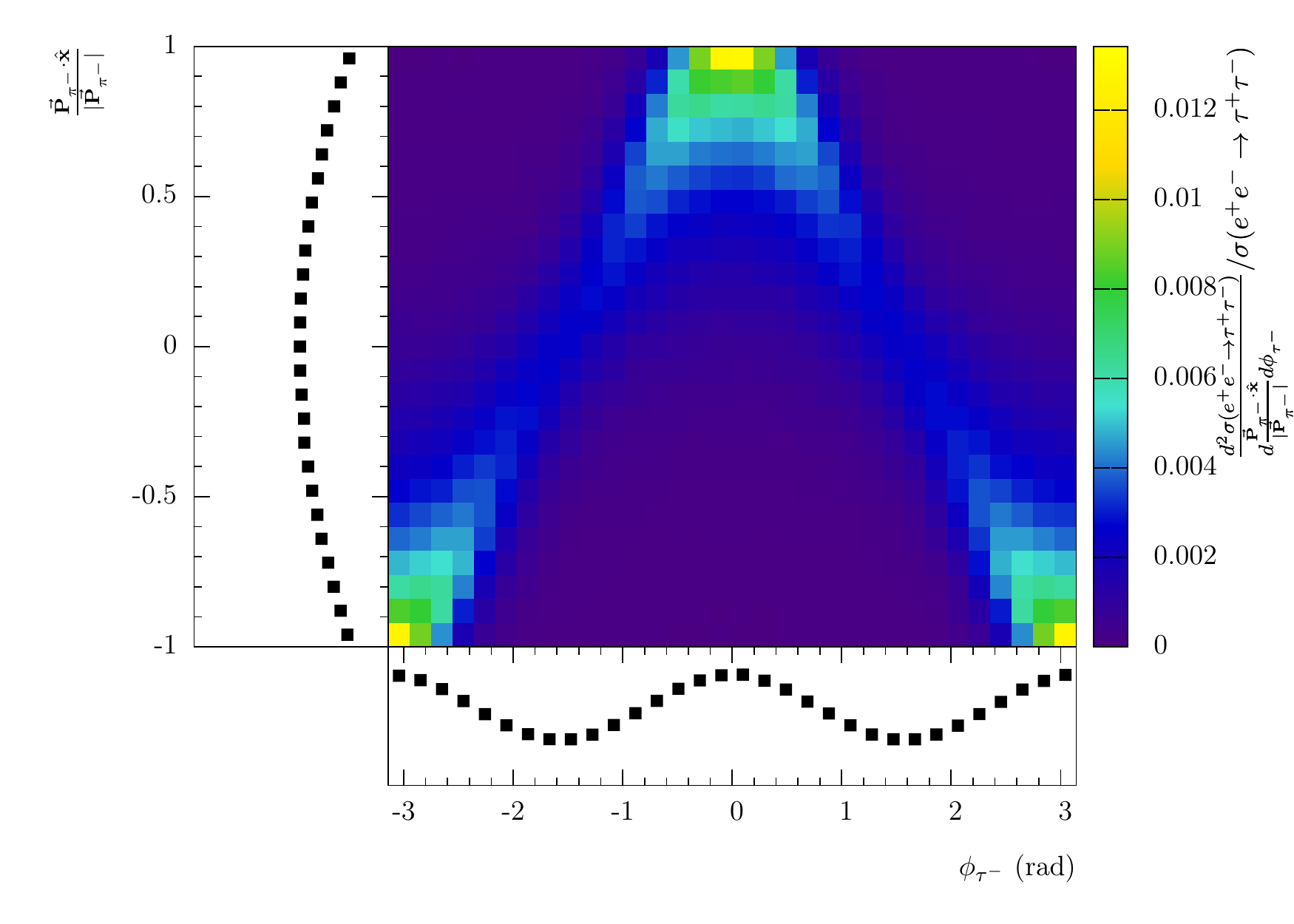}
  }
  \caption{An overlay of the missing transverse momentum (left) in a 100\% longitudinal polarized sample and for 100\% 
transversely polarized (along the x-axis) $e^{+}$ and $e^{-}$ sample along with the $cos(\theta_{\pi^{-}})$ $\phi_{\pi^{-}}$  
dependence of the cross-section weighted by $\frac{\vec{P}_{\pi^{-}}\cdot \hat{\mathbf{x}}}{|\vec{P}_{\pi^{-}|}}$. To simplify the
interpretation of these diagrams, they have been simulated at Born level.  \label{fig:TransvereObs}}
\end{center}
\end{figure*}

At relativistic energies, the total cross-sections are found to be independent of the transverse polarization of the initial state polarization which is consistent with known ultra-relativistic  
spin sum calculations and trace theorems for Dirac spinors in which the cross-section depends only on the longitudinal polarization independent of the basis \cite{Mandl:1985bg,Halzen:1984mc}.
This is in contrast to the non-relativistic energy,
 $|P|<<E$\footnote{This is achieved by artificially taking the limit where the mass of the initial-state electron and positron approaches $\to\sim\sqrt{s}/2$.}, where mass effects are dominant, 
the cross-section depends on the spin and is independent of the momentum direction, which is consistent with the non-relativistic quantum mechanics expectation for spin $1/2$ particles. 
 Moreover, the longitudinal spin observables in $\tau$ decays are independent 
of the transverse spin component as would be expected from the wave-function description where the
transverse spin is formed from a linear combinations of equal spin up and spin down helicity states. This is illustrated with the longitudinal spin observables $M_{\pi\pi}$ \cite{Pierzchala:2001gc} 
in Figure \ref{fig:LPolar} for $cos(\theta_{\pi})>0$ and  $cos(\theta_{\pi})<0$.  However, there are several observables related to the transverse polarization which are found to have a non-negligible 
impact on experimentally accessible distributions. In particular, the transverse polarization in the initial state leptons can generate polarization dependant oscillations in the $\phi$ distributions
missing transverse energy as seen in Figures \ref{fig:LPolar} and \ref{fig:TransvereObs}. This is related to the spin correlations in the decay of the $\tau$ lepton in which the 
neutrinos are aligned along the polarization axis 
in the transverse plane (Figure \ref{fig:TransvereObs}). For comparison, oscillations in the transverse momentum distributions from the p-wave emission of scalar mesons are expected in 
$e^{+}e^{-}\to\Upsilon(4s)\to B^{+}B^{-}/B^{0}\bar{B}^{0}$ production when there is a transverse spin component in the initial state.

\section{Conclusions}

A modified spin algorithm, which includes spin correlations, has been implemented in {\tt ee$\in$MC} to allow for the inclusion of an arbitrary initial spin
configuration of the electron-positron pair suitable for the proposed BELLE-II polarization upgrade to SuperKEKB \cite{USBELLEIIGROUP:2022qro,Roney:2021pwz}. 
The algorithm can be updated on an event-by-event basis to allow for a distribution of spin states including  tails caused by mis-alignment or spin relaxation distributions and to allow 
for the inclusion of changing detector conditions. The spin algorithm was implemented using both an extension to the modified Altarelli-Parisi Density Function to include the transverse spin of the 
initial state electron and positron as well as implementing a change of basis for the initial state wave-function of the colliding electron and positron. Both approaches yielded consistent results.  
The predictions are found to be consistent with theoretical expectations for the total cross-section and longitudinal observables.

\section*{Acknowledgement}
GCC Version 4.8.5 was used for compilation and the plots are generated using the external program GNUPlot  \cite{gnuplot4.2}.

\footnotesize
\bibliography{paper}
\normalsize


\noindent

\noindent

\end{document}